\documentclass[twocolumn,amsmath,amssymb,prl,superscriptaddress,a4paper]{revtex4}

\usepackage{graphicx}% Include figure files
\usepackage{dcolumn}% Align table columns on decimal point
\usepackage{bm}% bold math
\usepackage{subfigure}
\usepackage{amsmath}
\usepackage{amsfonts}
\usepackage{amssymb}
\usepackage{subfigure}
\usepackage[nice]{nicefrac}
\usepackage[tight]{units}

\begin{document}

\title{Phase Measurement of Resonant Two-Photon Ionization in Helium}

\author{M.~Swoboda}
\affiliation{Department of Physics, Lund University, P.O. Box 118, 22100 Lund, Sweden}

\author{T.~Fordell}
\affiliation{Department of Physics, Lund University, P.O. Box 118, 22100 Lund, Sweden}

\author{K.~Kl\"under}
\affiliation{Department of Physics, Lund University, P.O. Box 118, 22100 Lund, Sweden}

\author{J.~M.~Dahlstr\"om}
\affiliation{Department of Physics, Lund University, P.O. Box 118, 22100 Lund, Sweden}

\author{M.~Miranda}
\affiliation{Department of Physics, Lund University, P.O. Box 118, 22100 Lund, Sweden}
\affiliation{Departamento de F\'isica, Universidade do Porto, Rua do Campo Alegre 687, 4169-007 Porto, Portugal}

\author{C.~Buth}
\affiliation{Department of Physics and Astronomy, Louisiana State
University, Baton Rouge, Louisiana 70803, USA}
\affiliation{PULSE Institute, SLAC National Accelerator Laboratory, Menlo Park, California 94025, USA}

\author{K.~J.~Schafer}
\affiliation{Department of Physics and Astronomy, Louisiana State
University, Baton Rouge, Louisiana 70803, USA}
\affiliation{PULSE Institute, SLAC National Accelerator Laboratory, Menlo Park, California 94025, USA}

\author{J.~Mauritsson}
\affiliation{Department of Physics, Lund University, P.O. Box 118, 22100 Lund, Sweden}

\author{A.~L'Huillier}
\email{anne.lhuillier@fysik.lth.se}
\homepage{http://www.atto.fysik.lth.se}
\affiliation{Department of Physics, Lund University, P.O. Box 118, 22100 Lund, Sweden}

\author{M.~Gisselbrecht}
\affiliation{Department of Physics, Lund University, P.O. Box 118, 22100 Lund, Sweden}

%\date{\today}

%%
%%---------------------------------------------
%%
\begin{abstract}
We study resonant two-color two-photon ionization of Helium via the 1s3p
$^1P_1$ state. The first color is the 15$^\mathrm{th}$ harmonic of a tunable titanium sapphire laser, while the second color is the fundamental laser radiation. Our method uses phase-locked
high-order harmonics to determine the {\it phase} of the two-photon
process by interferometry. The measurement of the two-photon ionization
phase variation as a function of detuning from the resonance and intensity
of the dressing field allows us to determine the intensity dependence of
the transition energy.

\end{abstract}
%%
%%---------------------------------------------
%%
\maketitle
\begin{figure}
	\includegraphics[width=0.43\textwidth]{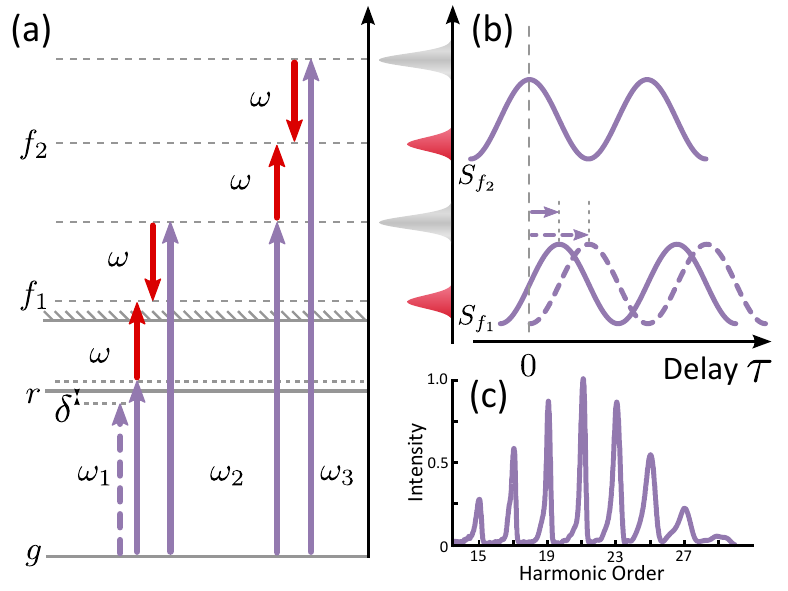}
\caption{(color online) a) Schematic diagram illustrating the phase measurement of R2PI. The dashed and solid $\omega_1$ lines represent two excitation energies on either side of the resonance. The photoelectron peaks used in the measurement are $S_{f_1}$ and $S_{f_2}$. b) Illustration of modulated sideband signals $S_{f_1}$ and $S_{f_2}$. Two $S_{f_1}$ modulations are indicated, corresponding to the two excitation energies in a).  c) Experimental harmonic spectrum used in the measurements.}
\label{principle}
\end{figure}

%%
%%---------------------------------------------
%%Intro
Multicolor resonant ionization is at the heart of numerous and diverse applications in fundamental and applied sciences. Examples are studies of very high Rydberg states \cite{OsterwalderPRL1999}, investigations of biomolecules \cite{RobertsonPCCP2001} and  specific selection of radioactive species \cite{KosterSAB2003}. In the simplest scheme, resonantly-enhanced two-photon ionization (R2PI) occurs via the absorption of two photons, generally of different colors, one tunable ($\omega_1$) used to scan across a resonant state ($r$), and the second ($\omega$) ionizing from the excited state. In traditional R2PI, the {\it yield} of the produced ion species is recorded as a function of laser wavelength, and the position and shape of the observed resonance provides information on the underlying electronic and rovibrational structures. These studies rely on spectroscopic information using narrow-bandwidth lasers, which do not allow any temporal resolution. Here, we present an ultrafast time-resolved-technique to retrieve also the {\it phase} of R2PI when sweeping through the resonance. We demonstrate it by studying R2PI of He via the 1s3p $^1$P$_1$ state which lies 23.087\,eV above the ground state.\\
\indent The basic principle of our experiment is illustrated in Fig. \ref{principle}. We study the interference between two pathways to the same ionized final state ($f_1$), one through the resonance with absorption of two photons with frequency $\omega_1$ and $\omega$, and the second through a continuum path, using a third color ($\omega_2$), involving absorption of a photon with frequency $\omega_2$ and emission of a photon with frequency $\omega$. The phase of the R2PI is encoded in the modulation of the photoelectron signal $S_{f_1}$ as a function of the delay $\tau$ between the ($\omega_1$,$\omega_2$) fields and the $\omega$ field [Fig. \ref{principle} (b)]. When the energy of the exciting radiation $\omega_1$, and thus the detuning from the resonance is changed, the phase variation of the resonant transition leads to a measurable shift of the $S_{f_1}$ oscillation. This phase shift needs to be referenced against another modulation $S_{f_2}$ that is independent of the resonance and thus providing a {\it clock} to our measurement. A process providing an independent modulation requires a fourth color ($\omega_3$) and involves another final state ($f_2$) (see Fig. \ref{principle}).\\
\indent An essential requirement for our measurement is the use of phase-locked radiation fields with commensurate frequencies, and a temporal precision better than the periodicity of the interference signal, in our case 1.3\,fs. Another requirement, is a high spectral resolution for the excitation of a narrow resonance. These requirements can be simultaneously fulfilled by using the high-order harmonic frequency combs produced when an intense laser field interacts with a gas of atoms or molecules \cite{FerrayJPB1988}. As is now well understood \cite{KrausePRL1992,CorkumPRL1993}, harmonics arise due to interferences between attosecond pulses produced by tunnel ionization, acceleration of the created wave packet in the field  and recombination back to the ground state at each half cycle of the laser field. The spectral width of the individual harmonics is thus related to the number of attosecond pulses, and decreases as the laser pulse duration increases \cite{BrandiPRL2003}. In this process, a comb of phase-locked harmonics of odd order is generated. 

In the present work, we use high-order harmonics to study two-color photoionization of He via the 1s3p $^1$P$_1$ state \cite{RanitovicNJP2010,HaberPRA2009}. In contrast to the ``reconstruction of attosecond bursts by interference of two-photon transition'' (RABITT) technique, used to determine the pulse duration of attosecond pulses \cite{VeniardPRA1996,PaulScience2001} and similarly to previous work performed in Ne \cite{VarjuLP2005} and N$_2$ \cite{HaesslerPRA2009}, we eliminate the influence of the temporal characteristics of the attosecond pulses to concentrate on the influence of the atomic properties. We study the R2PI phase as a function of detuning from the resonance, by varying the fundamental wavelength (around 805 nm) or alternatively by increasing the fundamental intensity. We apply these measurements to determine the intensity dependence of the energy of the 1s$^2~\rightarrow$~1s3p transition, and interpret the results using theoretical calculations consisting of solving the time-dependent Schr\"odinger equation (TDSE) in conditions close to the experimental ones \cite{SchaferPRL1997}.

% Setup
Our experiments were performed with a 1-kHz 35-fs 4-mJ Titanium-Sapphire laser system. An acousto-optic programmable dispersive filter (DAZZLER) was used to change the central wavelength between 802.5 and 809.3\,nm, while maintaining the spectral width of the amplified pulses approximately equal to 25 nm.  
High-order harmonics were generated in a pulsed Ar gas cell, filtered using a spatial aperture and a 200-nm thick Al thin film \cite{LopezMartensPRL2005}, and focused by a toroidal mirror into a vacuum chamber containing an effusive He gas jet. A magnetic bottle electron spectrometer (MBES) allowed us to record and analyze in energy the ejected electrons. Part of the laser field was extracted before the generation of harmonics, and recombined downstream collinearly with the harmonics, after a variable time delay that could be adjusted with sub-100-as precision \cite{VarjuLP2005}.

A comb of about seven phase-locked harmonic fields [Fig. \ref{principle} c)], corresponding in the time domain to a train of attosecond pulses of 260 as duration, was thus sent into the interaction chamber together with the dressing field at frequency $\omega$ with an adjustable phase $\varphi$ (or time $\tau=\varphi/\omega$) delay. In addition, a halfwave plate and polarizer in the dressing IR field arm allowed precise control of the pulse energy and therefore the intensity in the interaction region of the MBES. The detuning was determined from $\delta=15hc/\lambda_0 -E_{3p}$, where $E_{3p}$ is 23.087\,eV and $\lambda_0$ is the barycenter of the fundamental frequency spectrum, shifted to the blue by $\delta\lambda \simeq 3.5$\,nm to account for the blueshift from free electrons in the generation gas \cite{WahlstromPRA1993,GaardePRA2006}. The dressing laser intensity was determined by measuring the energy shifts of the photoelectron peaks of harmonics 17 to 23 in the presence of the laser field, which is simply equal to the ponderomotive energy $U_p \approx  6.0 I\,\mathrm{eV}$ where the intensity $I$ is in units of $\mathrm{10^{14}Wcm^{-2}}$ for a laser wavelength of 800\,nm \cite{FreemanPRL1987,BurnettJPB1993}.

\begin{figure}
\includegraphics[width=0.43\textwidth]{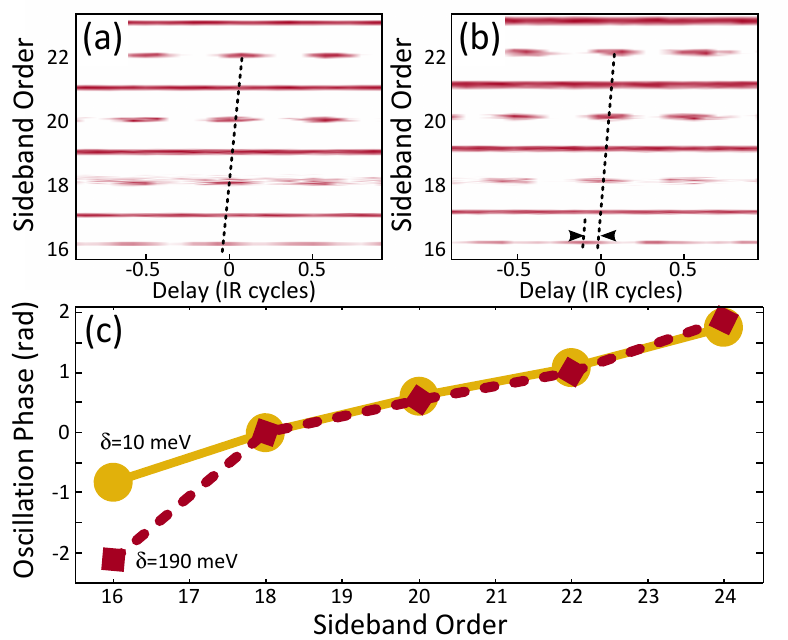}
\caption{(color online) Electron spectra as a function of delay for detunings $\delta=$10\,meV(a) and $\delta=$190\,meV(b). The oscillation of the 16$^{\mathrm{th}}$ sideband depends on the detuning while the others do not (see dashed line) (c) Phase of the oscillations of the sideband peaks in light orange for (a) and dark red for (b). The two results have been superposed at sideband 18.}
\label{rabitt}
\end{figure}

%Experimental results
Figure \ref{rabitt} (a,b) presents electron spectra as a function of delay between the harmonics and the 
dressing field, obtained for two different detunings. Electrons are observed at energies corresponding to one-photon absorption of the harmonics (from the 17$^\mathrm{th}$) and at ``sideband'' energies due to two-photon ionization processes, which we label by the corresponding net number of infrared photons (16, 18 etc.). These sideband peaks strongly oscillate with the delay at a frequency equal to $2\omega$. The oscillations of sidebands 18, 20, 22 and 24 do not depend on the detuning, while sideband 16 is strongly affected by it. A Fourier transform of the sideband signal over about 10 fs (four cycles) allows us to determine the relative phases of the sideband oscillations with a precision of $0.1\,\mathrm{rad}$. The phases are plotted in Fig.~\ref{rabitt} (c) for the two cases shown in (a,b). 

The relationship between the R2PI phase and the experimental results in Fig.~\ref{rabitt} can be understood within second-order perturbation theory \cite{VeniardPRA1996,VarjuLP2005}. Using the notation from Fig.~\ref{principle}, the photoelectron signal $S_{f_1}$ can be expressed as 
\begin{equation}
S_{f_1}=\left|a_1^a+a_2^e\right|^2
\label{eq:sf1}
\end{equation}
where $a_1^a$ and $a_2^e$ are the two-photon probability amplitudes with the superscript $a$ or $e$ referring to an absorption or emission of an $\omega$ photon and with the subscript $1$ or $2$ referring to absorption of an $\omega_1$ or $\omega_2$ photon. Introducing $\varphi_{1}$ and $\varphi_{2}$ as the phases of the radiation fields, as well as $\varphi_{1}^a$ and $\varphi_{2}^e$ as the phase terms involved in the two-photon transitions, Eq.~(\ref{eq:sf1}) becomes
\begin{eqnarray}
S_{f_1} & = & \left||a_1^a|e^{i\varphi_{1}^a+i\varphi_1+i\varphi}+|a_2^e|e^{i\varphi_{2}^a+i\varphi_2-i\varphi}\right|^2\\\nonumber
      & = & \left|a_1^a\right|^2\!+\!\left|a_2^e\right|^2\!+\!2\left|a_1^aa_2^e\right|\cos\left({\varphi_1^a\!-\!\varphi_2^e\!+\!2\varphi\!+\!\varphi_1\!-\!\varphi_2}\right) .
\end{eqnarray}
The cosine term leads to the modulation of the signal observed in the experiment. In general, the phase terms involved do not depend much on the photon energies. In two-photon ionization via a resonant state, however, the phase ($\varphi_1^a$) changes by $\pi$ across the resonance. The study of the variation of $\varphi_1^a$ as a function of detuning $\delta$ provides interesting information on the two-photon ionization process, e.g. on the relative importance of resonant and nonresonant contributions, AC Stark shift of the resonant state, depending on the spectral characteristics of the XUV and laser fields.  

\begin{figure}
\includegraphics[width=0.43\textwidth]{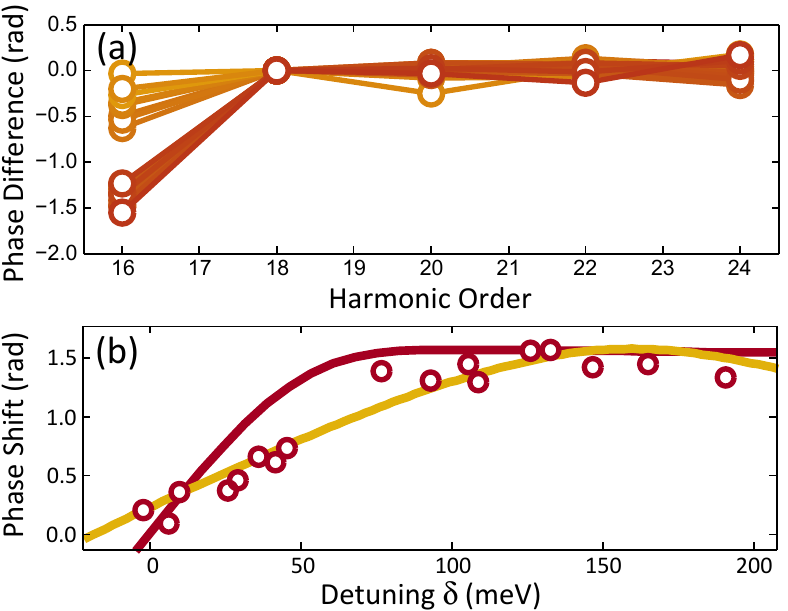}
\caption{(color online) a) Measured sideband phases corrected for the attosecond group delay and normalized at zero for sideband 18. Different detunings are indicated by the color code (going from 11\,meV below the resonance (dark red) to 190\,meV above the resonance (light orange)) b) Measurements (circles) of the R2PI phase as a function of detuning. The dark red line indicates results of a simple perturbative model while the light orange line shows results of simulations based on solving the TDSE.}
\label{wavelength}
\end{figure}

The variation of $\varphi_{1}^a$ with the detuning can be experimentally obtained from $S_{f_1}(\varphi)$ provided the other phase terms $\varphi_{2}^e$, $\varphi_1$, $\varphi_2$ do not depend on $\delta$ and provided the phase delay $\varphi$ is known in absolute value, which is generally not the case. $S_{f_1}(\varphi,\delta)$ is therefore referenced against $S_{f_2}(\varphi)$, assuming that the phase terms involved, $\varphi_{2}^a$ and $\varphi_{3}^e$, are independent of the detuning and thus removing the need of knowledge of the absolute $\varphi$. When changing $\delta$, the laser intensity used to generate the harmonics varies slightly, leading to a (small) variation of the group delay of the attosecond pulses and thus of $\varphi_1-\varphi_2$. We take this effect into account by assuming a linear group delay \cite{MairesseScience2003}, which we experimentally determine using higher-order sidebands. Its contribution is then subtracted from the measured phases and the phase of sideband 18 is set to zero for all detunings. The results are presented in Fig. \ref{wavelength}(a). As expected, the phases corresponding to all sidebands except the 16$^{\mathrm{th}}$ are almost superposed to each other and show no influence of detuning. 

Figure \ref{wavelength}(b) shows the variation of the R2PI phase as a function of detuning. We can tune only over half the resonance since for lower (negative detunings) sideband 16 moves progressively below the ionization threshold, thus making our phase measurement inaccurate. We also compare our measurements with the results of two different calculations (solid lines): The dark red line is obtained by a simple perturbative model \cite{DudovichPRL2002}, only considering the resonant state. Gaussian envelopes were used for the IR and XUV pulses with FWHMs of 30\,fs and 10\,fs, respectively. The light orange curve shows the result of calculations performed by numerically integrating the TDSE in the single active electron approximation \cite{SchaferPRL1997} in conditions close to the experiment. We use a He pseudo-potential with the energy of the 1s3p state equal to 23.039\,eV. The result is therefore shifted by 40\,meV for comparison with the experiment. The result shown in Fig. \ref{wavelength}(b) agrees very well with the experiment, thus confirming our detuning calibration.

\begin{figure}
\includegraphics[width=0.43\textwidth]{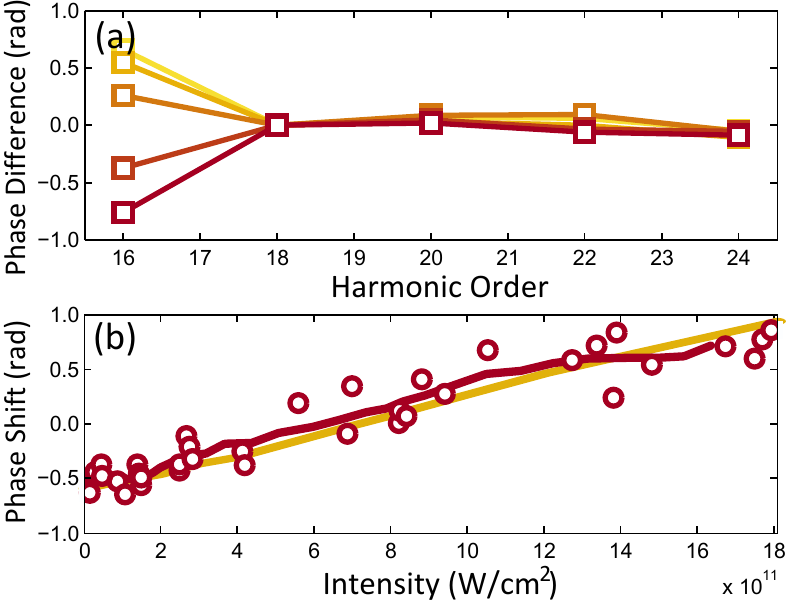}
\caption{(color online) Intensity dependence of the R2PI phase. (a) Harmonic phase differences at dressing intensities from $0.1$ (dark red) to $2.2\times 10^{12}\,\mathrm{Wcm^{-2}}$ (light orange), with attosecond chirp correction and normalization at sideband 18. (b) R2PI phase (circles) as a function of intensity, with a 6-point moving average (dark red line) and TDSE (light orange).}
\label{intensity}
\end{figure}

In order to investigate how the 1s3p resonance behaves in a laser field, we measured the dependence of the R2PI phase on the dressing laser intensity. Figure \ref{intensity}(a) shows the R2PI phase determined similarly to Figure \ref{wavelength}(a) but keeping the wavelength constant at 805.5\,nm and gradually increasing the dressing intensity. We verified that even at the highest intensity, higher-order multi-photon transitions were still negligible \cite{SwobodaLP2009}, thus not affecting our phase determination. Increasing the intensity from 0.1 to $1.8 \times\mathrm{10^{12}\,W/cm^2}$, the R2PI phase varies from  -0.7 to 0.9 radians. Figure \ref{intensity}(b) presents the intensity dependence of all of the measured phases (circles). We find an almost linear increase of the phase with intensity, as indicated by the dark red curve obtained by averaging, with a saturation at around $1.3\times 10^{12}\,\mathrm{W/cm^2}$, due to the suppression of R2PI when part of the two-photon excitation bandwidth moves partly below the ionization threshold. The light orange line obtained by TDSE calculations agrees well with our measurements.

\begin{figure}
\includegraphics[width=0.43\textwidth]{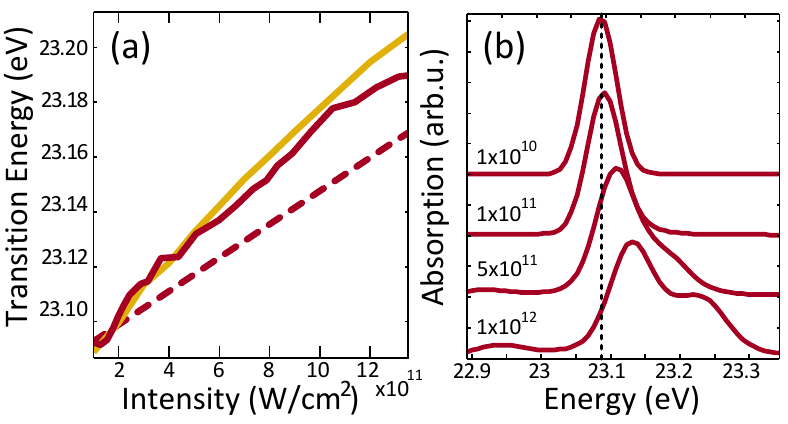}
\caption{(color online) (a) Measured transition energy of the 1s3p state. Experimental results (solid dark red), compared with $\Delta E_{1s^2}+U_p$ (dashed) and TDSE calculation (solid light orange). (b) TDSE calculation of XUV absorption for three different intensities with 50\,meV resolution. The position of the 1s3p state is indicated by the dashed line.}
\label{fig5}
\end{figure}

Combining our previous phase measurements as a function of detuning for a fixed (low) intensity and as a function of intensity (for a fixed detuning) allows us to determine the intensity-dependence of the $1s^2\rightarrow1s3p$ transition energy. Both experimental (dark red solid) and TDSE (light orange solid) results are shown in Fig. \ref{fig5}(a). The dashed line is equal to $\Delta E_{1s^2} + U_p$, representing the variation of the transition energy if the 1s3p state was moving as a high-lying Rydberg state, following $U_p$ \cite{BurnettJPB1993}. The AC Stark shift of the fundamental state $\Delta E_{1s^2}$ is very small, equal to $-0.3 I\mathrm{eV}$ where the intensity $I$ is in units of $\mathrm{10^{14}Wcm^{-2}}$ \cite{PerryPRL1989,RudolphPRL1991} so that $\Delta E_{1s^2}+U_p\approx U_p$. We find that the measured transition energy increases about 40\% more rapidly with the laser intensity than $U_p$, up to the saturation at $1.3\times 10^{12}\,\mathrm{Wcm^{-2}}$.

To better understand this faster than ponderomotive shift, we have calculated the XUV absorption cross-section for helium in the presence of an 800\,nm field by numerically solving the TDSE as a function of both XUV wavelength and laser intensity [Figure \ref{fig5}(b)]. Using an XUV bandwidth of 50\,meV or smaller we find that beyond $1\times10^{11}\,\mathrm{Wcm^{-2}}$, the 3p resonance has at least two components the higher of which shifts significantly faster than the ponderomotive energy. With the experimental XUV bandwidth (150 meV), however, the different components cannot be resolved. As a result, we observe shifts exceeding $E_{1s^2}+U_p$.  Experimentally, the predicted structure in the 3p resonance could be observed using longer fundamental laser pulses, leading to spectrally narrower harmonic peaks.

In conclusion, we have shown how well-characterized phase-locked high-order harmonics can be used to measure the phase of R2PI and we have applied it to the determination of the AC Stark shift of the 1s3p~$^1$P$_1$ state. Although our resolution was unsufficient to detect the splitting of the excited state, we observed a non-trivial, faster than ponderomotive, AC~Stark shift. Our method, here demonstrated in He, could be applied to the study of numerous resonant or quasi-resonant processes in atoms and molecules. 

We thank A.\,Maquet and R.\,Ta\"ieb for fruitful suggestions at the beginning of this work. This research was supported by the Marie Curie Intra-European Fellowship ATTOCO, the Marie Curie Early Stage Training Site (MAXLAS), the European Research Council (ALMA), the Knut and Alice Wallenberg Foundation and the Swedish Research Council. Funding at LSU is provided by the National Science Foundation through grant numbers PHY-0449235 and PHY-0701372.

%\bibliographystyle{aip}
%\bibliography{//Afbackup/ATTO/Library/Ref_lib}

\hyphenation{Post-Script Sprin-ger}

\end{document}